# Wall Shear Stress Analysis in Stenosed Carotid Arteries with Different Shapes of Plaque

Ruchika Bhatia
Student
Electrical, Electronics and Communication Engineering Department,
The Northcap University
Gurgaon, 122017
Haryana, India

Sharda Vashisth, PhD
Associate Professor
Electrical, Electronics and Communication Engineering Department
The Northcap University
Gurgaon, 122017
Haryana, India

Renu Saini
Student
Electrical, Electronics and Communication Engineering Department,
The Northcap University
Gurgaon, 122017
Haryana, India

## ABSTRACT
Atherosclerosis is a disease caused due to formation of plaque into the artery. Increase in plaque affects the wall shear stress. The present study is performed to calculate wall shear stress in different geometries of stenosed carotid artery.

A 2D model of different geometries is generated using CFD for Non- Newtonian model. After this WSS of different geometries of stenosed arteries is calculated and compared. Wall Shear Stress (WSS) of carotid arteries with smooth plaque, irregular plaque, cosine plaque and artery with blood clot is calculated.

It is found that with increase of plaque in common carotid artery WSS increases. Irregular plaque causes highest WSS. Wall Shear Stress of opposite walls of carotid artery is compared where one wall is having blood clot into it and other one is healthy.

## 1. INTRODUCTION
Heart attack is one of the leading causes of death worldwide. It occurs due to blockage into the artery. This blockage into the blood vessels is known as Atherosclerosis. If the blockage is in coronary artery heart attack occurs and if blockage is in cerebral artery stroke occurs. Carotid artery is one of the arteries which are prone to the blockage. Carotid artery is situated on both sides of the neck. The main function of this artery is to supply oxygenated blood and nutrients to face, neck and head. The supply of nutrients and blood is cut off. The blockage into the artery is known as plaque. Artery becomes hard due to plaque. The situation becomes worse if the plaque rupture. This rupturing of plaque triggers the blood to accumulate on the site of plaque. This gives rise to formation of blood clot. The blockage into the arteries increases the Wall Shear Stress. Wall Shear Stress is one of the important characteristic of blood flow and has strong link in formation of plaque.

Initially, research on Atherosclerosis was conducted using experimental techniques by visualizing flow through a tube partially obstructed. Flow characteristics were calculated by varying shape of plaque and its complexity, site and fluid flowing through the vessel. Dye tracing method was used to gain more accurate and deep information considering flow recirculation and flow separation. Laser Doppler anemometry provided visualization of nature of blood flow without influencing it. But this method has limitations and is complex. It is difficult to calculate characteristics like wall shear stress by these experimental methods. Piezoelectric sensor was also used to acquire carotid artery [3]. Experimental methods do not gave accurate results close to the wall region. Recently, computer simulations used provide best results. Even the complex processes can be easily simulated by using these non invasive methods.

Computational techniques have been developed to simulate arteries as they are easy to use and economical. Computational methods provide a platform to simulate an artery using software with conditions as close to as real scenario. Blood flow conditions are taken similar to the real flow. These methods help to obtain the relationship between blood flow and shape of stenos artery. Computational values are compared with non invasive methods. It makes the analysis and research somewhat easy.

A study [21] concluded that flow behavior depends upon complexity, length of plaque, axisymmetric and asymmetric conditions and Reynolds numbers by studying the steady flow through stenosed arteries. An artery was also simulated which was stenosed assuming a Newtonian fluid [15]. Power Law and Quemada non-Newtonian models were used to study the pulsatile nature of non-Newtonian blood flow through the arteries which are having plaque into them. [4]. This study deduced that flows with a high pulsatile nature tend to generate very complex flow patterns. A study was employed to simulate symmetric carotid artery to calculate Wall shear stress. Blood was assumed to be Newtonian. It was concluded that WSS is directly proportional to length and height of the plaque [15].

## 2. METHOD
Computational fluid dynamics is a method used to simulate the blood flow and solve equations of the fluid flow. It divides the whole geometry into different number of cells. This study uses Ansys software which gives fast and accurate results. All the geometries of blocked carotid artery are shown in Fig.1, 2, 3 and 4. The block diagram which shows all the steps taken for this study is shown in Fig. 5.

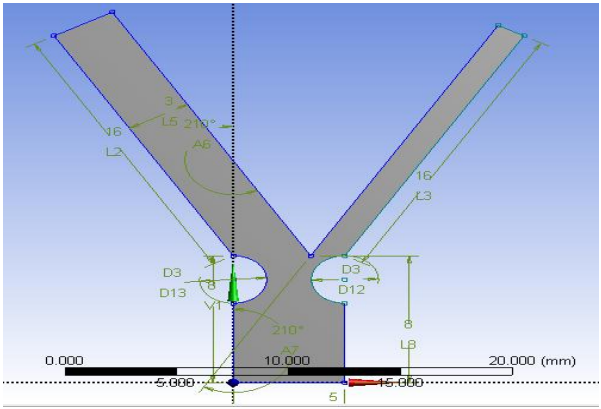

**Fig.1: Carotid artery having smooth plaque**

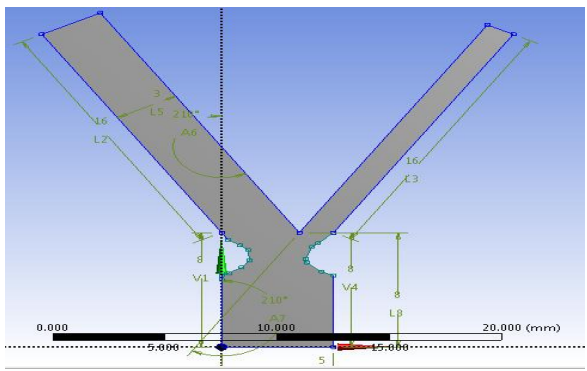

**Fig.2: Carotid artery having cosine plaque**

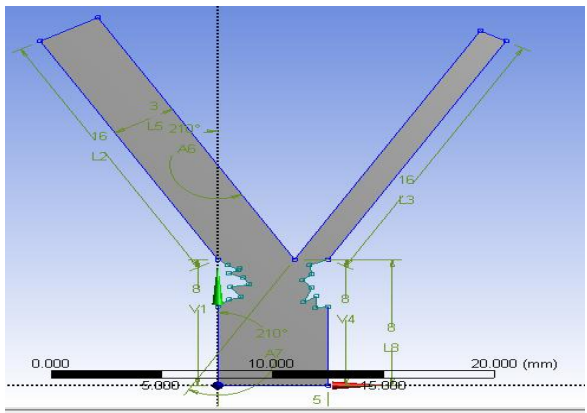

**Fig.3: Carotid artery having irregular plaque**

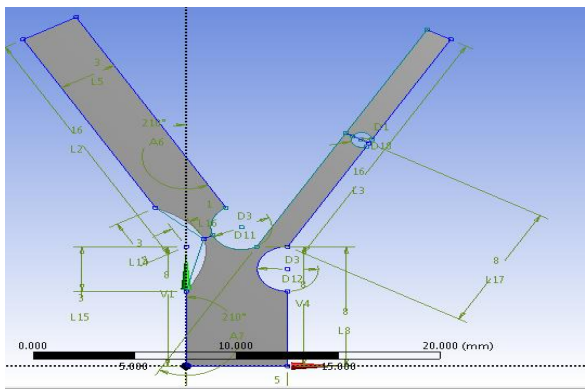

**Fig. 4: Carotid artery having clot**

In this study model used to represent blood flow is Carreau Non-Newtonian model. In this paper three different geometries of plaque in carotid artery are compared. The three types of plaque used are smooth, cosine and irregular. Also in this paper carotid artery having blood clot is generated. The wall of carotid artery having clot is compared with opposite wall of same carotid artery having no clot. The Carreau [5, 12] Non-Newtonian model used in this present study is represented by the following equation:

$$\mu = \mu_\infty + (\mu_0 - \mu_\infty)[1+(\lambda\gamma)^2]^{(n-1)/2} \quad \ldots\ldots\ldots \quad (1)$$

where time constant, $\lambda = 3.313$ s, $n = 0.3568$, zero strain viscosity (i.e. resting viscosity),

$\mu_0 = 0.56$ P and infinite strain viscosity, $\mu_\infty = 0.0345$ P.

Density of blood is taken as 1060Kg/m^3. Gauge outlet pressure is 0 Pa for stenosed artery. The artery is assumed to be cylindrical and symmetrical. 2D steady flow and geometry is used. Coefficient of viscosity is taken as 0.001Kg/(m-s). Blood flow is pulsatile and cyclic.

WSS is inversely proportional to the diameter of artery and directly proportional to the viscosity of blood flow. Plaque formed into the carotid artery decreases the area of artery and hence increases the WSS. It is given by the equation:

$$\tau = \frac{F}{A} \quad \ldots\ldots\ldots\ldots\ldots\ldots\ldots\ldots\ldots \quad (2)$$

where $\tau$ is WSS and F is force applied and A is area of cross section.

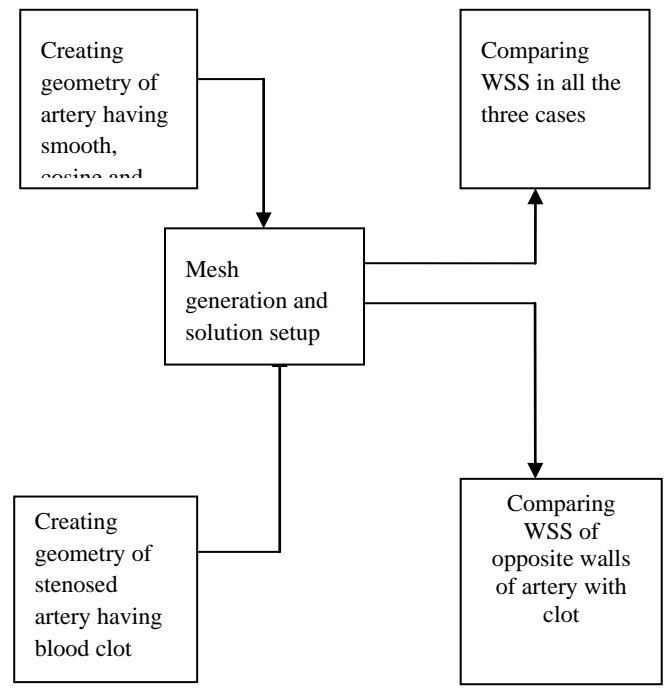

**Fig. 5: Representation of practical block diagram**

## 3. RESULTS

Wall Shear Stress is one of the important characteristic of blood. Wall Shear Stress of carotid artery is calculated in case of all three types of plaque. It is found that Wall Shear Stress in case of irregular plaque is highest. WSS in case of cosine and smooth plaque is approximately same as shown in Fig.8. Irregular plaque has irregular troughs and peaks. Cosine and smooth plaque is having similar average Wall Shear Stress around the wall selected. The Wall Shear Stress is calculated for the wall of artery which is shown in Fig.6, 7 by the polyline.

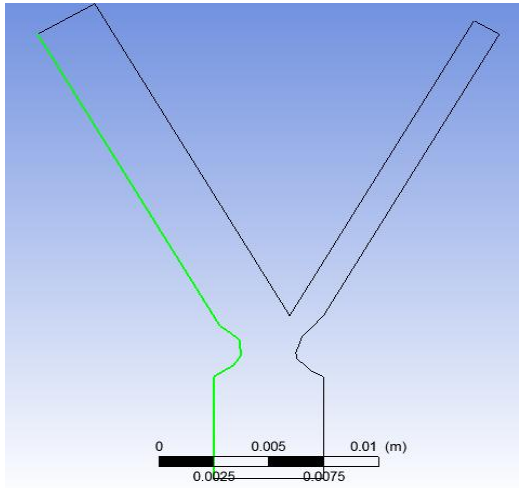

**Fig. 6: Representation of artery wall along which WSS due to cosine plaque has to be calculated**

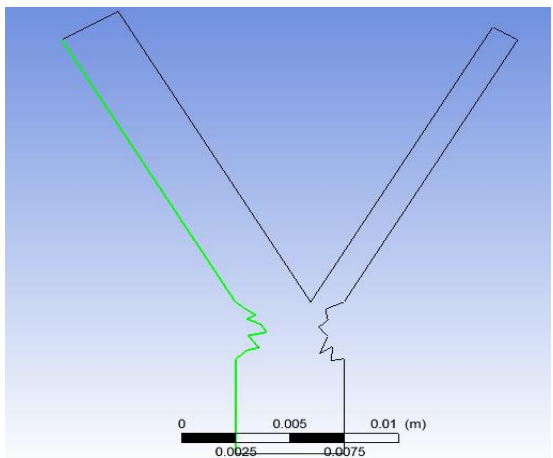

**Fig. 7: Representation of artery wall along which WSS due to irregular plaque has to be calculated**

**Table 1. Average WSS in different stenosed arteries**

| Type of Plaque | Average Wall Shear Stress |
|---|---|
| Irregular | 1.74 |
| Cosine | 1.5 |
| Smooth | 1.6 |

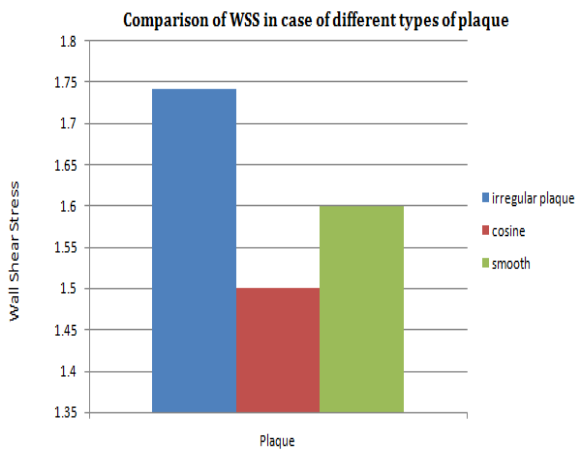

**Fig.8: Comparison of WSS on the basis of type of plaque**

In case of artery having clot, two opposite walls of artery are compared. One wall is having clot in it and other one is healthy which are represented in Fig.9.

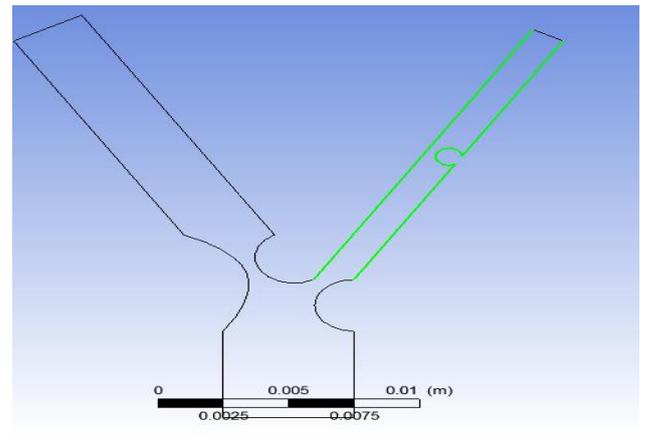

**Fig.9: Wall arteries whose comparison is done**

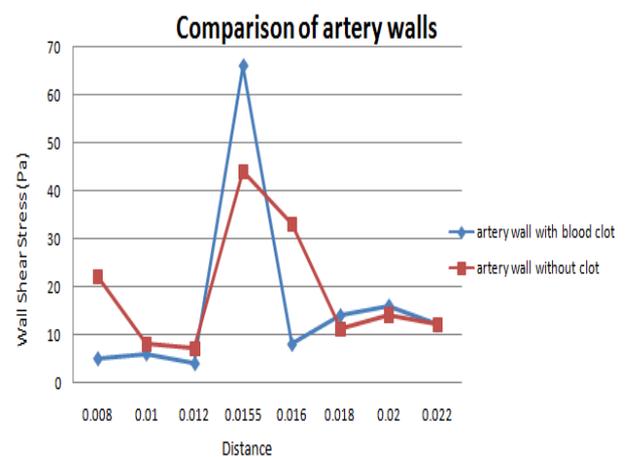

**Fig.10: Comparison of WSS in artery walls with and without clot**

In the beginning WSS of the two walls is same. The first point at 0.008 which is 8mm is different for both the walls. This is because both the walls are having different starting points. At 0.0155 WSS is highest as this point is at the centre of clot. It gives maximum WSS value for both the walls as shown in Fig.10. After that at point 0.016 which is point just at the end of clot, WSS is less as compared to healthy artery. WSS is same for both the arteries afterwards.

**Table 2. WSS in case of artery having blood clot**

| Distance(mm) | Wall Shear Stress(Pa) |
|---|---|
| 0.008 | 22 |
| 0.01 | 08 |
| 0.012 | 07 |
| 0.0155 | 44 |
| 0.016 | 33 |
| 0.018 | 11 |
| 0.02 | 14 |
| 0.022 | 12 |

## 4. CONCLUSION

Atherosclerosis is one of the major causes of death in western world. It is increasing rapidly. Heart attack and Stroke are also caused due to Atherosclerosis.

We studied four different cases of blockage in this study. Three different types of plaque smooth, cosine and irregular are simulated. Fourth case is of an artery which is having blood clot in it. Artery having irregular plaque is found to be having more Wall Shear Stress as compared to arteries having cosine and smooth plaque. Wall Shear Stress in case of smooth and cosine plaque is approximately similar. It is concluded that irregular surface increases Wall Shear Stress in the walls of the artery.

Wall Shear Stress remains same for the two opposite walls of artery. But if clot occurs Wall Shear Stress increases first and then decreases to become same as WSS in artery without clot. It is found that Wall Shear Stress starts to increase with beginning of the clot. As clot area increases Wall shear Stress also increases. Wall Shear Stress is found to be changing with the area of clot. At the centre of clot, Wall Shear Stress is highest. After that it starts to decrease with decrease in the area of clot. Then it becomes exactly as the Wall Shear Stress of opposite wall of artery having no clot.

WSS depends upon plaque. It varies with area of the blockage. It is highest at centre and decreases afterwards.